\documentclass[proceedings]{JHEP3}
\usepackage{amsfonts}
\usepackage{amsmath}
\usepackage{epsfig,multicol}

\newbox\mybox

\newcommand\fverb{\setbox\mybox=\hbox\bgroup\verb}
\newcommand\fverbdo{\egroup\medskip\noindent\fbox{\unhbox\mybox}\ }
\newcommand\fverbit{\egroup\item[\fbox{\unhbox\mybox}]}
\conference{$\mathcal{PT}$ invariant complex $E_8$ root spaces}
\abstract{We provide a construction procedure for complex root spaces invariant under antilinear transformations, which may be applied to any Coxeter group. The procedure is based on the factorisation of
a chosen element of the Coxeter group into two factors. Each of the factors constitutes an 
involution and may therefore be deformed in an antilinear fashion.
Having the importance of the $E_{8}$-Coxeter group in mind, such as
underlying a particular perturbation of the Ising model and the
fact that for it no solution could be found previously, we exemplify the procedure for this particular case. As a concrete application of this construction we propose new generalisations of Calogero-Moser-Sutherland models and affine Toda field theories based on the invariant complex root spaces and deformed complex simple roots, respectively.}
\title{$\mathcal{PT}$ invariant complex $E_8$ root spaces}
\author{Andreas Fring and Monique Smith \\
Centre for Mathematical Science, City University London,\\
Northampton Square, London EC1V 0HB, UK\\
E-mail: a.fring@city.ac.uk , abbc991@city.ac.uk}

\input{tcilatex}

\begin{document}

\section{Introduction}

It is known for more than twenty years that symmetries based on the $E_{8}$%
-Lie group or $E_{8}$-Coxeter (Weyl) group are known to be important in the
context of 1+1 dimensional integrable models. In a field theoretical context
A.B. Zamolodchikov \cite{Zamolodchikov:1989fp} found in 1989 that the
conformal field theory with central charge $c=1/2$ perturbed by the primary
field $\phi _{(1,2)}$ of conformal weight $\Delta =1/6$ gives rise to an
affine Toda field theory with an $E_{8}$-mass spectrum. On the lattice side
this field theory was identified to correspond to the Ising model in a
magnetic field. Remarkably, the first experimental evidence supporting these
theoretical findings were reported only very recently in \cite{Coldea:2010zz}%
.

Furthermore, it is known that the Ising model may be perturbed by a complex
field \cite{Guenter1,CastroAlvaredo:2009vq} and still describe a meaningful
physical system, despite of being related to a non-Hermitian Hamiltonian.
This is related to the fact that the non-Hermitian Hamiltonian possess the
property of being $\mathcal{PT}$-symmetric in a wider sense, meaning that it
remains invariant under a simultaneous parity transformation $\mathcal{P}$
and time reversal $\mathcal{T}$. Strictly speaking the Hamiltonian remains
invariant under the more general transformation of an antilinear involutory
map of which $\mathcal{PT}$-symmetry is only one example. Then by an
observation of Wigner\ \cite{EW}, made already fifty years ago, the
eigenvalues of the Hamiltonian, or any other operator with that symmetry
property, are guaranteed to be real when in addition also their
eigenfunctions possess this symmetry. More recently many new physically
meaningful models have been constructed and properties of older models could
be explained consistently exploiting this feature, for recent reviews see
e.g. \cite{Bender:1998ke,Benderrev,Rev3}. While this type of representation
is usually very simple to verify for single particle Hamiltonians it is less
obviously identified in multi-particle systems or field theories. Often the
symmetry is only evident after a suitable change of variables or even a full
separation of variables \cite{Milos,FZ}. Since many of such type of models
are formulated generically in terms of root systems, as for instance
Calogero-Moser-Sutherland models \cite{OPPP} or Toda field theories \cite%
{Wilson,DIO}, with the dynamical variables or fields lying in the dual
space, the possibility to deform directly these roots was explored recently 
\cite{FZ,FringSmith}. This approach allows to deal with a huge class of
models in a very systematic manner as it provides a well defined scheme when
based on the roots rather than on a deformation of the canonical variables
or fields.

The general logic followed was to identify first an element $w$ in the Weyl
(Coxeter) group $w\in \mathcal{W}$ with the involutory property $w^{2}=%
\mathbb{I}$, view it as the analogue of the $\mathcal{P}$-operator and
subsequently deform it in an antilinear fashion. The most obvious candidates
to take are simple Weyl reflections. However, it was shown in \cite%
{FringSmith} that root spaces with the desired properties based on this
identification can only be constructed for groups of rank 2. The explicit
solutions for the groups $A_{2}$, $G_{2}$ and $B_{2}$ can be found in \cite%
{FZ} and \cite{Assis:2009gt}, respectively. In \cite{FringSmith} we
identified the analogue of the $\mathcal{P}$-operator with either of the two
factors $\sigma _{+}$ or $\sigma _{-}$ of the Coxeter element $\sigma $ in
the form $\sigma =\sigma _{-}\sigma _{+}$ or the longest element $w_{0}$ of
the Weyl group. In both cases we were able to construct explicitly the
invariant complex root spaces for a large number of groups. However, we
could also show that in many cases an explicit solution does either not
exist based on the identifications used or leads only to trivial solutions.

In particular, no non-trivial deformation of the $E_{8}$-root system which
remain invariant under antilinear transformations was found. Motivated in
addition by the above mentioned importance the $E_{8}$-root systems play,
the main purpose of this manuscript is to provide such a deformation.
However, our procedure is very general and may in principle be applied to
any element in any group.

In comparison with previous approaches we select here factorisations of an
element in the Coxeter group, say $\tilde{\sigma}$, of order $\tilde{h}$
less than the Coxeter number $h$, i.e. $\tilde{\sigma}^{\tilde{h}}=\mathbb{I}
$. We factorise them similarly as the usual Coxeter element based on the
bi-colouration of the Dynkin diagram and by construction each of the two
factors are then involutory maps, since all subfactors are commuting Weyl
reflections being involutions themselves. We identify them as the analogue
of the parity transformation and deform them to build up an antilinear
involution. A reduced complex root space is then constructed from the orbits
of these elements containing $\ell =~$rank$~\mathcal{W}\times \tilde{h}$
roots instead of the $\ell \times h$ roots, which result when generated from
the usual Coxeter element. By construction these root systems remain
invariant under the action of each of the deformed factors of the chosen
element when certain conditions hold.

With the above mentioned motivation in mind, one may then employ the
deformed simple roots to define complex versions of $E_{8}$-affine Toda
field theories or the entire deformed root space to formulate new complex
versions of Calogero-Moser-Sutherland models. We report the properties of
these models elsewhere \cite{FSprep}.

\section{From factorised Weyl group elements to invariant complex rootspaces}

Let us now briefly recall the main aim of the method of construction
proposed so far. We use the notation of \cite{FringSmith} and refer to it
and references therein for parts of the definitions used. The aim is to
construct complex extended root systems $\tilde{\Delta}(\varepsilon )$ which
remain invariant under a newly defined antilinear involutary map. The
standard real roots $\alpha _{i}\in \Delta \subset \mathbb{R}^{n}$ are
sought to be represented in a complex space depending on some deformation
parameter $\varepsilon \in \mathbb{R}$ as $\tilde{\alpha}_{i}(\varepsilon
)\in \tilde{\Delta}(\varepsilon )\subset $ $\mathbb{R}^{n}\oplus \imath 
\mathbb{R}^{n}$. For this purpose we define a linear deformation map 
\begin{equation}
\delta :~\Delta \rightarrow \tilde{\Delta}(\varepsilon ),\qquad \alpha
\mapsto \tilde{\alpha}=\theta _{\varepsilon }\alpha ,  \label{d}
\end{equation}%
relating simple roots $\alpha $ and deformed simple roots $\tilde{\alpha}$
in a linear fashion via the constant deformation matrix $\theta
_{\varepsilon }$. Subsequently we seek an antilinear involutory map $\omega $
which leaves this root space invariant%
\begin{equation}
\omega :\tilde{\Delta}(\varepsilon )\rightarrow \tilde{\Delta}(\varepsilon
),\qquad \tilde{\alpha}\mapsto \omega \tilde{\alpha},  \label{omega}
\end{equation}%
this means the map satisfies $\omega :\tilde{\alpha}=\mu _{1}\alpha _{1}+\mu
_{2}\alpha _{2}\mapsto \mu _{1}^{\ast }\omega \alpha _{1}+\mu _{2}^{\ast
}\omega \alpha _{2}$ for $\mu _{1}$, $\mu _{2}\in \mathbb{C}$ and $\omega
^{2}=\mathbb{I}$. Clearly there are many possibilities to achieve this.

As already mentioned, what has been investigated this far is to take simple
Weyl reflections as candidates for $\omega $, which works successfully for
rank 2 groups, the two factors $\sigma _{\pm }$ of the Coxeter element or
the longest element $w_{0}$ of the Weyl group. What has not been explored
this far is to take different types of elements in $\mathcal{W}$ as starting
points. Here we will indicate the general procedure and work out explicitly
the concrete $E_{8}$-example. A more systematic solution procedure for other
cases will be provided elsewhere \cite{FSprep}.

We will start with an arbitrary element of the Weyl group $\tilde{\sigma}\in 
\mathcal{W}$. This means the element can by definition always be expressed
as a product over simple Weyl reflections $\tilde{\sigma}=\prod \sigma _{i}$%
. Due to the fact that Weyl reflections do not commute there are various
ways to represent elements in the same similarity class. We will therefore
convert this element always into a factorised form of the following type%
\begin{equation}
\tilde{\sigma}=\tilde{\sigma}_{-}\tilde{\sigma}_{+}\quad \ \ \ \ \ \ \text{%
with \ \ \ }\tilde{\sigma}_{\pm }:=\prod\limits_{i\in \tilde{V}_{\pm
}}\sigma _{i},
\end{equation}%
in close analogy to the factorisation of the Coxeter element $\sigma =\sigma
_{-}\sigma _{+}$ as explained in \cite{FringSmith} and references therein.
The sets $V_{\pm }$ are defined via the bi-colouration, meaning that the
roots are separated into two sets of disjoint roots on the Dynkin diagram.
However, the products do not extend over all possible elements, i.e. $\tilde{%
V}_{\pm }\subset V_{\pm }$ and therefore we may think of these elements as 
\begin{equation}
\tilde{\sigma}_{\pm }:=\sigma _{\pm }\prod\limits_{i\in \tilde{V}_{\pm
}^{\prime }}\sigma _{i}
\end{equation}%
for some values $j$, with $\tilde{V}_{\pm }^{\prime }\cup \tilde{V}_{\pm
}=V_{\pm }$, by recalling $[\sigma _{i},\sigma _{j}]=0$ for $i,j\in V_{+}$
or $i,j\in V_{-}$ and $\sigma _{i}^{2}=\mathbb{I}$. This ensures that we
maintain the crucial property $\tilde{\sigma}_{\pm }^{2}=\mathbb{I}$ and
thus we select $\tilde{\sigma}_{-}$ or $\tilde{\sigma}_{+}$ as a potential
candidates for the analogue of the $\mathcal{P}$-operator which we seek to
deform in an antilinear fashion to construct the map $\omega $ introduced in
(\ref{omega}). This is achieved by defining the antilinear deformations of
the factors of the modified Coxeter element as%
\begin{equation}
\tilde{\sigma}_{\pm }^{\varepsilon }:=\theta _{\varepsilon }\tilde{\sigma}%
_{\pm }\theta _{\varepsilon }^{-1}=\tau \tilde{\sigma}_{\pm }
\end{equation}%
with $\tau $ acting as a complex conjugation and $\theta _{\varepsilon }$
being the deformation matrix intoduced in (\ref{omega}). By similar
reasoning as in \cite{FringSmith} we find that the properties to be
satisfied by $\theta _{\varepsilon }$ are 
\begin{equation}
\theta _{\varepsilon }^{\ast }\tilde{\sigma}_{\pm }=\tilde{\sigma}_{\pm
}\theta _{\varepsilon },\quad \left[ \tilde{\sigma},\theta _{\varepsilon }%
\right] =0,\quad \theta _{\varepsilon }^{\ast }=\theta _{\varepsilon
}^{-1},\quad \det \theta _{\varepsilon }=\pm 1\quad \text{and\quad }%
\lim_{\varepsilon \rightarrow 0}\theta _{\varepsilon }=\mathbb{I}\text{.}
\label{const}
\end{equation}%
These equations will be enough to determine the deformed simple roots $%
\tilde{\alpha}$. Before defining the entire root space associated to $\tilde{%
\sigma}^{\varepsilon }=\tilde{\sigma}_{-}^{\varepsilon }\tilde{\sigma}%
_{+}^{\varepsilon }$ we introduce the root space $\tilde{\Delta}$ associated
to $\tilde{\sigma}.$ We require for this the values $c_{i}=\pm 1$ assigned
to the vertices of the Coxeter graphs, in such a way that no two vertices
with the same values are linked together. Using then the quantity $\gamma
_{i}=c_{i}\alpha _{i}$ for a representant similarly as in the undeformed
case, we define a "reduced" Coxeter orbit as 
\begin{equation}
\tilde{\Omega}_{i}:=\left\{ \gamma _{i},\tilde{\sigma}\gamma _{i},\tilde{%
\sigma}^{2}\gamma _{i},\ldots ,\tilde{\sigma}^{\tilde{h}-1}\gamma
_{i}\right\} ,
\end{equation}%
and the entire reduced root space as 
\begin{equation}
\tilde{\Delta}:=\bigcup\limits_{i=1}^{\ell }\tilde{\Omega}_{i}\subset \Delta
_{\mathcal{W}}.  \label{omega2}
\end{equation}%
The length of the orbits $\tilde{\Omega}_{i}$ will naturally be reduced
because the order of the element $\tilde{\sigma}$ will be smaller than the
Coxeter number $h$%
\begin{equation}
\tilde{\sigma}^{\tilde{h}}=\mathbb{I},\qquad \ \ \ \ \ \ \ \text{with }%
\tilde{h}\leq h.
\end{equation}%
This means the total number of roots in $\tilde{\Delta}$ is $\ell \tilde{h}$
rather than $\ell h$ as in the case of $\Delta _{\mathcal{W}}$. Since $\left[
\tilde{\sigma},\theta _{\varepsilon }\right] =0$, the deformed orbits and
root spaces are isomorphic to the undeformed ones and we may define deformed
reduced orbits and a deformed rootspace as 
\begin{equation}
\tilde{\Omega}_{i}^{\varepsilon }:=\theta _{\varepsilon }\tilde{\Omega}%
_{i}\qquad \text{and\qquad }\tilde{\Delta}(\varepsilon ):=\theta
_{\varepsilon }\tilde{\Delta},
\end{equation}%
respectively. Crucial to our construction is that the deformed root space $%
\tilde{\Delta}(\varepsilon )$ remains invariant under the antilinear
involutory transformation $\tilde{\sigma}_{\pm }^{\varepsilon }:\tilde{\Delta%
}(\varepsilon )\rightarrow \tilde{\Delta}(\varepsilon )$. This follows from
the argument 
\begin{equation}
\tilde{\sigma}_{\pm }^{\varepsilon }:\tilde{\Delta}(\varepsilon )\rightarrow
\theta _{\varepsilon }\tilde{\sigma}_{\pm }\theta _{\varepsilon }^{-1}\tilde{%
\Delta}(\varepsilon )=\theta _{\varepsilon }\tilde{\sigma}_{\pm }\tilde{%
\Delta}=\theta _{\varepsilon }\tilde{\Delta}=\tilde{\Delta}(\varepsilon )
\label{a}
\end{equation}%
if and only if $\tilde{\sigma}_{\pm }\tilde{\Delta}=\tilde{\Delta}$. As the
root space is now reduced this might not be the case as $\tilde{\sigma}_{\pm
}$ could map a root into the complement of $\tilde{\Delta}$. However, we may
verify this explicity on a case-by-case basis.

\section{Invariant complex $E_{8}$-root spaces}

Our conventions for the labelling of the $E_{8}$-roots are depicted in the
Dynkin diagram below. They differ slightly from the one previously used \cite%
{FringSmith}, but have the advantage that neither two roots labelled by odd
or even numbers are connected, which allows for compact notation.

\unitlength=0.6500000pt 
\begin{picture}(370.00,107.58)(180.00,0.00)
\put(170.00,38.00){\makebox(0.00,0.00){$E_8:$}}

\put(485.00,52.00){\makebox(0.00,0.00){$\alpha_8$}}
\put(445.00,52.00){\makebox(0.00,0.00){$\alpha_7$}}
\put(405.00,52.00){\makebox(0.00,0.00){$\alpha_6$}}
\put(365.00,52.00){\makebox(0.00,0.00){$\alpha_5$}}
\put(335.00,52.00){\makebox(0.00,0.00){$\alpha_4$}}
\put(285.00,52.00){\makebox(0.00,0.00){${\alpha}_3$}}
\put(325.00,90.00){\makebox(0.00,0.00){${\alpha}_1$}}
\put(245.00,52.00){\makebox(0.00,0.00){${\alpha}_2$}}

\put(450.00,38.00){\line(1,0){30.00}}
\put(410.00,38.00){\line(1,0){30.00}}
\put(370.00,38.00){\line(1,0){30.00}}
\put(330.00,38.00){\line(1,0){30.00}}
\put(325.33,72.67){\line(0,-1){31.00}}
\put(290.00,38.00){\line(1,0){30.00}}
\put(250.00,38.00){\line(1,0){30.00}}

\put(325.00,78.00){\circle*{10.00}}
\put(245.00,38.00){\circle*{10.00}}
\put(325.00,38.00){\circle*{10.00}}
\put(365.00,38.00){\circle*{10.00}}
\put(405.00,38.00){\circle*{10.00}}
\put(445.00,38.00){\circle*{10.00}}
\put(485.00,38.00){\circle*{10.00}}
\put(285.00,38.00){\circle*{10.00}}
%%%%%%%%%%%%%%%%
\end{picture}

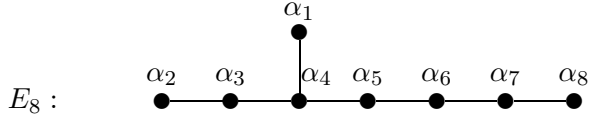
\smallskip \noindent {\small Figure 1: Dynkin diagram indicating the
conventions of the labelling for the simple $E_{8}$-roots.}

Let us now illustrate the procedure described above by selecting first of
all an element of the Weyl group, for instance%
\begin{equation}
\tilde{\sigma}=\sigma _{1}\sigma _{-}\sigma _{+}=\sigma _{3}\sigma
_{5}\sigma _{7}\sigma _{2}\sigma _{4}\sigma _{6}\sigma _{8}.
\end{equation}%
Acting on the vector $\vec{\alpha}=\{\alpha _{1},\alpha _{2},\alpha
_{3},\alpha _{4},\alpha _{5},\alpha _{6},\alpha _{7},\alpha _{8}\}$ we can
represent this element as%
\begin{equation}
\tilde{\sigma}=\left( 
\begin{array}{rrrrrrrr}
1 & 0 & 0 & 1 & 0 & 0 & 0 & 0 \\ 
0 & 0 & 1 & 1 & 0 & 0 & 0 & 0 \\ 
0 & -1 & -1 & -1 & 0 & 0 & 0 & 0 \\ 
0 & 1 & 1 & 1 & 1 & 1 & 0 & 0 \\ 
0 & 0 & 0 & -1 & -1 & -1 & 0 & 0 \\ 
0 & 0 & 0 & 1 & 1 & 1 & 1 & 1 \\ 
0 & 0 & 0 & 0 & 0 & -1 & -1 & -1 \\ 
0 & 0 & 0 & 0 & 0 & 1 & 1 & 0%
\end{array}%
\right) .  \label{1}
\end{equation}%
Simply by matrix multiplication we then find that the order of this element
is $\tilde{h}=8$. Using an Ansatz for the deformation matrix similar to the
one in \cite{FringSmith}%
\begin{equation}
\theta _{\varepsilon }=c_{0}\mathbb{I}+(1-c_{0})\tilde{\sigma}^{4}+i\sqrt{%
c_{0}^{2}-c_{0}}(\tilde{\sigma}^{2}-\tilde{\sigma}^{-2}),\qquad c_{0}\in 
\mathbb{R},  \label{2}
\end{equation}%
we find that the first four constraints in (\ref{const}) are satisfied. The
ususal choice $c_{0}=\cosh \varepsilon $ ensures that we recover the
undeformed case in the limit $\varepsilon \rightarrow 0$, that is the last
requirement in (\ref{const}). Explicitly the deformation matrix resulting
from (\ref{1}) and (\ref{2}) reads%
\begin{equation}
\theta _{\varepsilon }=\left( 
\begin{array}{cccccccc}
1 & \lambda _{0} & 2\lambda _{0}-i\kappa _{0} & 3-3c_{0} & 3\lambda
_{0}-i\kappa _{0} & 3-3c_{0} & 2\lambda _{0}-i\kappa _{0} & \lambda _{0} \\ 
0 & c_{0} & 0 & i\kappa _{0} & 2i\kappa _{0} & i\kappa _{0} & 0 & -\lambda
_{0} \\ 
0 & 0 & c_{0}-i\kappa _{0} & -2i\kappa _{0} & -2i\kappa _{0} & -2i\kappa _{0}
& -i\kappa _{0}-\lambda _{0} & 0 \\ 
0 & i\kappa _{0} & 2i\kappa _{0} & c_{0}+2i\kappa _{0} & 2i\kappa _{0} & 
2i\kappa _{0}-\lambda _{0} & 2i\kappa _{0} & i\kappa _{0} \\ 
0 & -2i\kappa _{0} & -2i\kappa _{0} & -2i\kappa _{0} & 2(c_{0}-i\kappa
_{0})-1 & -2i\kappa _{0} & -2i\kappa _{0} & -2i\kappa _{0} \\ 
0 & i\kappa _{0} & 2i\kappa _{0} & 2i\kappa _{0}-\lambda _{0} & 2i\kappa _{0}
& c_{0}+2i\kappa _{0} & 2i\kappa _{0} & i\kappa _{0} \\ 
0 & 0 & -\lambda _{0}-i\kappa _{0} & -2i\kappa _{0} & -2i\kappa _{0} & 
-2i\kappa _{0} & c_{0}-i\kappa _{0} & 0 \\ 
0 & -\lambda _{0} & 0 & i\kappa _{0} & 2i\kappa _{0} & i\kappa _{0} & 0 & 
c_{0}%
\end{array}%
\right) .
\end{equation}%
In order to achieve a more compact notation we introduced here the
abbreviations $\kappa _{0}=\sqrt{c_{0}^{2}-c_{0}}$ and $\lambda _{0}=1-c_{0}$%
. Therefore the deformed simple roots resulting from (\ref{d}) are%
\begin{eqnarray}
\tilde{\alpha}_{1} &=&\alpha _{1}+\lambda _{0}\left( \alpha _{2}+2\alpha
_{3}+3\alpha _{4}+3\alpha _{5}+3\alpha _{6}+2\alpha _{7}+\alpha _{8}\right)
-i\kappa _{0}\left( \alpha _{3}+\alpha _{5}+\alpha _{7}\right) , \\
\tilde{\alpha}_{2} &=&c_{0}\left( \alpha _{2}+\alpha _{8}\right) -\alpha
_{8}+i\kappa _{0}\left( \alpha _{4}+2\alpha _{5}+\alpha _{6}\right) , \\
\tilde{\alpha}_{3} &=&c_{0}\left( \alpha _{3}+\alpha _{7}\right) -\alpha
_{7}-i\kappa _{0}\left[ \alpha _{3}+2\left( \alpha _{4}+\alpha _{5}+\alpha
_{6}\right) +\alpha _{7}\right] , \\
\tilde{\alpha}_{4} &=&c_{0}\left( \alpha _{4}+\alpha _{6}\right) -\alpha
_{6}+i\kappa _{0}\left[ \alpha _{2}+2\left( \alpha _{3}+\alpha _{4}+\alpha
_{5}+\alpha _{6}+\alpha _{7}\right) +\alpha _{8}\right] , \\
\tilde{\alpha}_{5} &=&\left( 2c_{0}-1\right) \alpha _{5}-2i\kappa _{0}\left(
\alpha _{2}+\alpha _{3}+\alpha _{4}+\alpha _{5}+\alpha _{6}+\alpha
_{7}+\alpha _{8}\right) , \\
\tilde{\alpha}_{6} &=&c_{0}\alpha _{6}+\alpha _{4}\left( c_{0}+2i\kappa
_{0}-1\right) +i\kappa _{0}\left[ \alpha _{2}+2\left( \alpha _{3}+\alpha
_{5}+\alpha _{6}+\alpha _{7}\right) +\alpha _{8}\right] , \\
\tilde{\alpha}_{7} &=&c_{0}\alpha _{7}+(c_{0}-1)\alpha _{3}-i\kappa _{0}%
\left[ \alpha _{3}+2\left( \alpha _{4}+\alpha _{5}+\alpha _{6}\right)
+\alpha _{7}\right] , \\
\tilde{\alpha}_{8} &=&c_{0}\alpha _{8}-\lambda _{0}\alpha _{2}+i\kappa
_{0}\left( \alpha _{4}+2\alpha _{5}+\alpha _{6}\right) .
\end{eqnarray}%
To construct the invariant root space we compute the undeformed reduced root
space $\tilde{\Delta}$ from the orbits $\tilde{\Omega}_{i}$ for $1\leq i\leq
8$ and subsequently replace the undeformed roots in a one-to-one fashion by
their deformed counterparts. We evaluate%
\begin{equation*}
\begin{tabular}{l||c|c|c|c|c|c|c|c|}
$\tilde{\Delta}$ & $\alpha _{1}$ & $\alpha _{2}$ & $\alpha _{3}$ & $\alpha
_{4}$ & $\alpha _{5}$ & $\alpha _{6}$ & $\alpha _{7}$ & $\alpha _{8}$ \\ 
\hline\hline
$\tilde{\sigma}$ & {\small 1;4} & {\small 3;4} & {\small -2;3;4} & {\small %
2;3;4;5;6} & {\small -4;5;6} & {\small 4;5;6;7;8} & {\small -6;7;8} & 
{\small 6;7} \\ \hline
$\tilde{\sigma}^{2}$ & {\small 1;2;3;4}$^{2}${\small ;5;6} & {\small 5;6} & 
{\small \negthinspace \negthinspace -3;4;5;6\negthinspace \negthinspace } & 
{\small \negthinspace \negthinspace 3;4;5;6;7;8\negthinspace \negthinspace }
& {\small \negthinspace \negthinspace -2;3;4;5;6;7;8\negthinspace
\negthinspace } & {\small \negthinspace \negthinspace
2;3;4;5;6;7\negthinspace \negthinspace } & {\small \negthinspace
\negthinspace -4;5;6;7\negthinspace \negthinspace } & {\small 4;5} \\ \hline
$\tilde{\sigma}^{3}$ & {\small \negthinspace \negthinspace 1;2;3}$^{2}$%
{\small ;4}$^{3}${\small ;5}$^{2}${\small ;6}$^{2}${\small ;7;8\negthinspace
\negthinspace } & {\small 7;8} & {\small \negthinspace \negthinspace
-5;6;7;8\negthinspace \negthinspace } & {\small 5;6;7} & {\small -3;4;5;6;7}
& {\small 3;4;5} & {\small \negthinspace \negthinspace -2;3;4;5\negthinspace
\negthinspace } & {\small 2;3} \\ \hline
$\tilde{\sigma}^{4}$ & {\small \negthinspace \negthinspace 1;2;3}$^{2}$%
{\small ;4}$^{3}${\small ;5}$^{3}${\small ;6}$^{3}${\small ;7}$^{2}${\small %
;8\negthinspace \negthinspace } & {\small -8} & {\small -7} & {\small -6} & 
{\small -5} & {\small -4} & {\small -3} & {\small -2} \\ \hline
$\tilde{\sigma}^{5}$ & {\small \negthinspace \negthinspace 1;2;3}$^{2}$%
{\small ;4}$^{3}${\small ;5}$^{3}${\small ;6}$^{2}${\small ;7}$^{2}${\small %
;8\negthinspace \negthinspace } & {\small \negthinspace \negthinspace
-6;7\negthinspace \negthinspace } & {\small 6;7;8} & {\small -4;5;6;7;8} & 
{\small 4;5;6} & {\small -2;3;4;5;6} & {\small 2;3;4} & {\small %
\negthinspace \negthinspace -3;4\negthinspace \negthinspace } \\ \hline
$\tilde{\sigma}^{6}$ & {\small 1;2;3}$^{2}${\small ;4}$^{2}${\small ;5}$^{2}$%
{\small ;6;7} & {\small \negthinspace \negthinspace -4;5\negthinspace
\negthinspace } & {\small 4;5;6;7} & {\small \negthinspace \negthinspace
-2;3;4;5;6;7\negthinspace \negthinspace } & {\small \negthinspace
\negthinspace 2;3;4;5;6;7;8\negthinspace \negthinspace } & {\small %
\negthinspace \negthinspace -3;4;5;6;7;8\negthinspace \negthinspace } & 
{\small 3;4;5;6} & {\small \negthinspace \negthinspace -5;6\negthinspace
\negthinspace } \\ \hline
$\tilde{\sigma}^{7}$ & {\small 1;3;4;5} & {\small \negthinspace
\negthinspace -2;3\negthinspace \negthinspace } & {\small 2;3;4;5} & {\small %
-3;4;5} & {\small 3;4;5;6;7} & {\small -5;6;7} & {\small 5;6;7;8} & {\small %
\negthinspace \negthinspace -7;8\negthinspace \negthinspace }%
\end{tabular}%
\end{equation*}%
We report in this table the roots which emerge as a \ result of computing $%
\tilde{\sigma}^{n}(\alpha _{i})$ with $1\leq n\leq 7$ and $1\leq i\leq 8$,
where we indicate multiple occurrences by a power. Since these root are
either positive or negative it suffices to report the overall sign. For
instance we read off from the table that $\tilde{\sigma}^{3}(\alpha
_{1})=\alpha _{1}+\alpha _{2}+2\alpha _{3}+3\alpha _{4}+2\alpha _{5}+2\alpha
_{6}+\alpha _{7}+\alpha _{8}$ or $\tilde{\sigma}^{2}(\alpha _{3})=-\alpha
_{3}-\alpha _{4}-\alpha _{5}-\alpha _{6}$.

Next we compute the action of $\tilde{\sigma}_{\pm }$ on the simple roots.
We find%
\begin{equation}
\begin{array}{ll}
\tilde{\sigma}_{-}\alpha _{1}=\alpha _{1},~~ & \tilde{\sigma}_{-}\alpha
_{2}=\alpha _{2}+\alpha _{3}=\tilde{\sigma}^{3}\alpha _{8},~~ \\ 
\tilde{\sigma}_{-}\alpha _{3}=-\alpha _{3},~~ & \tilde{\sigma}_{-}\alpha
_{4}=\alpha _{3}+\alpha _{4}+\alpha _{5}=\tilde{\sigma}^{3}\alpha _{6}, \\ 
\tilde{\sigma}_{-}\alpha _{5}=-\alpha _{5},~~ & \tilde{\sigma}_{-}\alpha
_{6}=\alpha _{5}+\alpha _{6}+\alpha _{7}=\tilde{\sigma}^{3}\alpha _{4},~~ \\ 
\tilde{\sigma}_{-}\alpha _{7}=-\alpha _{7},~~~~~ & \tilde{\sigma}_{-}\alpha
_{8}=\alpha _{7}+\alpha _{8}=\tilde{\sigma}^{3}\alpha _{2},%
\end{array}
\label{a2}
\end{equation}%
and%
\begin{equation}
\begin{array}{ll}
\tilde{\sigma}_{+}\alpha _{1}=\alpha _{1}+\alpha _{4}=\tilde{\sigma}\alpha
_{1},~~ & \tilde{\sigma}_{+}\alpha _{2}=-\alpha _{2},~~ \\ 
\tilde{\sigma}_{+}\alpha _{3}=\alpha _{2}+\alpha _{3}+\alpha _{4}=\tilde{%
\sigma}^{5}\alpha _{7},~~ & \tilde{\sigma}_{+}\alpha _{4}=-\alpha _{4}, \\ 
\tilde{\sigma}_{+}\alpha _{5}=\alpha _{4}+\alpha _{5}+\alpha _{6}=\tilde{%
\sigma}^{5}\alpha _{5},~~ & \tilde{\sigma}_{+}\alpha _{6}=-\alpha _{6},~~ \\ 
\tilde{\sigma}_{+}\alpha _{7}=\alpha _{6}+\alpha _{7}+\alpha _{8}=\tilde{%
\sigma}^{5}\alpha _{3},~~ & \tilde{\sigma}_{+}\alpha _{8}=-\alpha _{8}.%
\end{array}
\label{a1}
\end{equation}%
From (\ref{a2}), (\ref{a1}) and the above table we observed that $\tilde{%
\sigma}_{\pm }\alpha _{i}\in \tilde{\Delta}$ for $1\leq i\leq 8$, such that $%
\tilde{\sigma}_{\pm }\tilde{\Delta}=\tilde{\Delta}$. Therefore replacing in
the above table simple roots by deformed simple roots, $\alpha _{i}\mapsto 
\tilde{\alpha}_{i}$ for $1\leq i\leq 8$, consitutes a complex root space $%
\tilde{\Delta}(\varepsilon )$ consisting of $64$ different complex roots
which remains invariant under the antilinear involutory transformations $%
\tilde{\sigma}_{\pm }^{\varepsilon }$. Note that in this particular case the
introduction of the representative $\gamma _{i}$, which is often needed to
avoid overcounting, is not essential.

In order to obtain non-trivial invariant root spaces with different amounts
of roots we may select elements $\tilde{\sigma}\in \mathcal{W}$ of other
order $\tilde{h}$. For instance we compute%
\begin{equation}
\begin{array}{ll}
\tilde{\sigma}=\sigma _{1}\sigma _{3}\sigma _{7}\sigma _{4}\sigma _{6}\sigma
_{8},~\ \ ~ & \text{with }\tilde{h}=4, \\ 
\tilde{\sigma}=\sigma _{1}\sigma _{3}\sigma _{5}\sigma _{7}\sigma _{4}\sigma
_{6}\sigma _{8},\qquad & \text{with }\tilde{h}=12, \\ 
\tilde{\sigma}=\sigma _{1}\sigma _{3}\sigma _{7}\sigma _{2}\sigma _{4}\sigma
_{6}\sigma _{8}, & \text{with }\tilde{h}=20, \\ 
\tilde{\sigma}=\sigma _{1}\sigma _{3}\sigma _{5}\sigma _{7}\sigma _{2}\sigma
_{4}\sigma _{8}, & \text{with }\tilde{h}=24.%
\end{array}%
\end{equation}%
Generalizing then the Ansatz (\ref{2}) to%
\begin{equation}
\theta _{\varepsilon }=c_{0}\mathbb{I}+(1-c_{0})\tilde{\sigma}^{\tilde{h}%
/4}+i\sqrt{c_{0}^{2}-c_{0}}(\tilde{\sigma}^{\tilde{h}/2}-\tilde{\sigma}^{-%
\tilde{h}/2}),
\end{equation}%
yields non-trivial deformation matrices satisfying the constraint (\ref%
{const}). In general we may try any element of the form%
\begin{equation}
\tilde{\sigma}=\prod\limits_{j\in \tilde{V}_{-}}\sigma
_{i}\prod\limits_{j\in \tilde{V}_{+}}\sigma _{i}=:\tilde{\sigma}_{-}\tilde{%
\sigma}_{+}  \label{si}
\end{equation}%
where the product might not extend over all four odd and four even roots in $%
V_{-}$ or $V_{+}$, respectively. It is clear that this creates a large
amount of possibilities. In \cite{FSprep} we report on how one may organise
these options systematically.

\section{Conclusions}

We have provided a general construction procedure for invariant root spaces
under antilinear involutory transformations. The starting point can be any
element in the Weyl group. Since by definition such elements consist of
products of Weyl reflections we may always bring it into a factorised form (%
\ref{si}), such that each of the factors is comprised of elements related to
simple roots which are not connected on the Dynkin diagram. Then each of the
factors $\tilde{\sigma}_{\pm }$ will be an involution, which we can identify
as an analogue to the $\mathcal{P}$-operator and subsequently we deform it
to introduce the antilinear maps $\tilde{\sigma}_{\pm }^{\varepsilon }$.
Solving the constraints (\ref{const}) we construct simple deformed roots.
The entire root space $\tilde{\Delta}(\varepsilon )$ may then be constructed
from the union of the reduced orbits $\tilde{\Omega}_{i}(\varepsilon )$. By
construction it remains invariant under the action of $\tilde{\sigma}_{\pm
}^{\varepsilon }$ if and only if $\tilde{\sigma}_{\pm }\tilde{\Delta}=\tilde{%
\Delta}$.

We may then employ these spaces to investigate new types of non-Hermitian
generalisation of Calogero models 
\begin{equation}
\mathcal{H}(p,q)=\frac{p^{2}}{2}+\frac{\omega ^{2}}{4}\sum_{\tilde{\alpha}%
\in \tilde{\Delta}(\varepsilon )}(\tilde{\alpha}\cdot q)^{2}+\sum_{\tilde{%
\alpha}\in \tilde{\Delta}(\varepsilon )}\frac{g_{\tilde{\alpha}}}{(\tilde{%
\alpha}\cdot q)^{2}},  \label{HC}
\end{equation}%
or the analogues of Calogero-Moser-Sutherland models when replacing the
rational potential by a trigonometric or elliptic one. We may also employ
only the deformed simple roots and investigate properties of generalised
versions of affine Toda field theories decribed by Lagrangians of the form 
\begin{equation}
\mathcal{L}=\frac{1}{2}\partial _{\mu }\phi \partial ^{\mu }\phi -\frac{m^{2}%
}{\beta ^{2}}\sum\limits_{i=0}^{\ell }n_{0}e^{\beta \tilde{\alpha}_{i}\cdot
\phi }.
\end{equation}

For the case of $E_{8}$ we have provided the explicit construction for the
rootspaces. Based on the conjecture that\ $E_{8}$ plays a crucial role in
the understanding and realisation of perturbations of the Ising model it is
conceivable that the complex deformed root systems serve to faciliate a
systematic study of non-Hermitian perturbations of the Ising model.
Generalisions to other types of spin chain and their scaled versions might
be based on generalisations to other Coxeter groups.

\medskip

\noindent \textbf{Acknowledgments:} MS is supported by EPSRC.

%%\bibliographystyle{phreport}
%%\bibliography{Ref}

\end{document}